\documentclass[aps,prl,twocolumn,superscriptaddress,notitlepage]{revtex4-2}
\pdfoutput=1
\usepackage{graphicx}
\usepackage{mathptmx, textcomp}
\usepackage[usenames]{xcolor}
\usepackage{dcolumn}
\usepackage{bm}
\usepackage{amssymb, amsmath}
\usepackage[colorlinks,urlcolor=blue,citecolor=blue,linkcolor=blue]{hyperref}
\usepackage{cleveref}
\usepackage{physics}
\usepackage{verbatim}
\usepackage{amsmath}
\usepackage[normalem]{ulem}
\usepackage{lineno}
\usepackage{float}
\usepackage{comment}
\usepackage[utf8]{inputenc}

\begin{document}

\title{Quantum beat spectroscopy of repulsive Bose polarons}

\author{A. \ M. \ Morgen}
\affiliation{Center for Complex Quantum Systems, Department of Physics and Astronomy, Aarhus University, Ny Munkegade 120, DK-8000 Aarhus C, Denmark}
\author{S.\ S.\ Balling}
\affiliation{Center for Complex Quantum Systems, Department of Physics and Astronomy, Aarhus University, Ny Munkegade 120, DK-8000 Aarhus C, Denmark}
\author{K.\ Knakkergaard Nielsen}
\affiliation{Max-Planck Institute for Quantum Optics, Hans-Kopfermann-Str. 1, D-85748 Garching, Germany}
\author{T.\ Pohl}
\affiliation{Center for Complex Quantum Systems, Department of Physics and Astronomy, Aarhus University, Ny Munkegade 120, DK-8000 Aarhus C, Denmark}
\affiliation{Institute for Theoretical Physics, Vienna University of Technology, Wiedner Hauptstraße 8-10, 1040 Vienna, Austria}
\author{G.\ M.\ Bruun}
\affiliation{Center for Complex Quantum Systems, Department of Physics and Astronomy, Aarhus University, Ny Munkegade 120, DK-8000 Aarhus C, Denmark}
\author{J. J. Arlt}
\affiliation{Center for Complex Quantum Systems, Department of Physics and Astronomy, Aarhus University, Ny Munkegade 120, DK-8000 Aarhus C, Denmark}
\date{\today}

\begin{abstract} The physics of impurities in a bosonic quantum environment is a paradigmatic and challenging many-body problem that remains to be understood in its full complexity. Here, this problem is investigated for impurities with strong repulsive interactions based on Ramsey interferometry in a quantum degenerate gas of $^{39}$K atoms. We observe an oscillatory signal that is consistent with a quantum beat between two co-existing coherent quasiparticle states: the attractive and repulsive polarons. The interferometric signal allows us to extract the polaron energies for a wide range of interaction strengths, complimenting earlier spectroscopic measurements. We furthermore identify several dynamical regimes towards the formation of the Bose polaron in good agreement with theory. Our results improve the understanding of quantum impurities interacting strongly with a bosonic environment, and demonstrate how quasiparticles as well as short-lived non-equilibrium many-body states can be probed using Ramsey interferometry.
\end{abstract}

\maketitle
As famously argued by Landau, a bare impurity particle smoothly evolves into a quasiparticle as the interaction with a surrounding quantum environment is adiabatically increased. Whereas the concept of quasiparticles was originally developed to understand the motion of electrons in solids~\cite{LandauPekar}, it now has a much broader scope and forms an incredibly successful platform for understanding strongly interacting quantum many-body systems~\cite{BaymPethick1991book,Nakano2020,Mannella2005, Baggioli2015,Lee2006}.

Ultracold quantum gas experiments have in recent years improved our understanding of quasiparticles, as they allow one to perform Landau's gedanken experiment by tuning the interaction strength between an impurity atom and a surrounding quantum gas. This has led to the observation of the Fermi polaron~\cite{Schirotzek2009,Kohstall2012,Scazza2017}, which is now well understood theoretically even for strong interactions~\cite{Massignan2014}. Furthermore, its non-equilibrium formation dynamics~\cite{cetina2015,cetina2016} and mediated interactions between two Fermi polarons~\cite{baroni2023mediated} have been observed. There is also strong experimental evidence for the Bose polaron with an energy~\cite{jorgensen2016,hu2016,Yan2020,ardila2019} and formation dynamics~\cite{Skou2020,Skou2022PRR} that matches theoretical predictions well for attractive interactions. For repulsive interactions, the spectroscopic measurements are generally complicated with broader signals due to damping, the possible presence of few-body bound states, and a many-body continuum. Except in the perturbative regime~\cite{Casteels2014,Christensen2015}, there are indeed many different theoretical predictions, reflecting the richness of the problem~\cite{Rath2013,Li2014,Levinsen2015,Shchadilova2016,Guenther2018,Yoshida2018,Drescher2020,Massignan2021,Levinsen2021,Schmidt2022,christianen2023phase,mostaan2023unified}.

\begin{figure}[t!]
    \centering
    \includegraphics[width=\columnwidth]{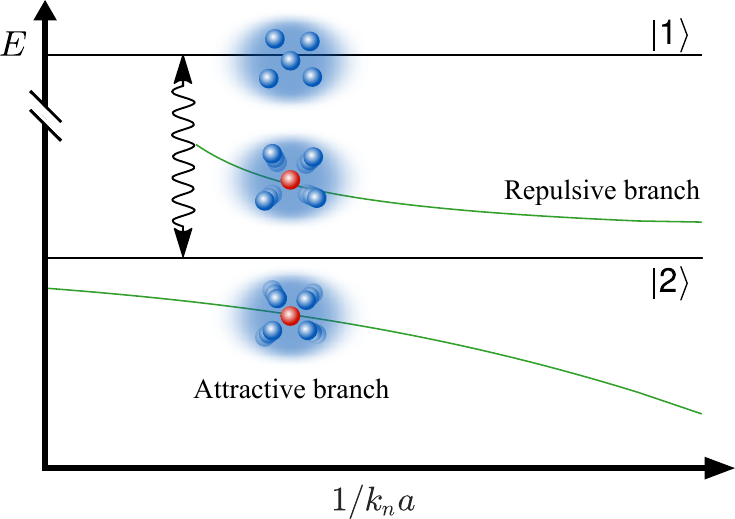}
    \caption{Illustration of the employed Ramsey interferometry method and a simplified energy landscape of the system as a function of inverse interaction strength $1/(k_n a)$. A first radio frequency (rf) pulse (black arrow) is applied to a Bose-Einstein condensate (BEC) in the initial electronic state $\ket{1}$ (blue dots), which generates a small admixture in the impurity state $\ket{2}$ (red dots). Its interaction with the background BEC results in attractive and repulsive Bose polaron branches (green lines). A second rf pulse closes the interferometric sequence. The presence of two polaronic branches leads to the observation of a beat signal in the amplitude of the oscillatory interferometer output.} 
    \label{fig:ind_fig}
\end{figure}

In this Letter, we use Ramsey interferometry to probe the non-equilibrium dynamics of impurities in a weakly interacting Bose-Einstein condensate (BEC) for repulsive interactions. The underlying method is illustrated in Fig.~\ref{fig:ind_fig}. We observe a fast initial decay and a revival in the amplitude of the oscillatory interferometric signal. This revival corresponds to a quantum beat, indicating the presence of two states, which we refer to as the repulsive and attractive branches. The interferometric measurements allow for a characterization of the energy of these branches as a function of interaction strength.  Based on the initial decay, several non-equilibrium dynamical regimes are identified and explained theoretically. These interferometric results provide valuable information on impurity dynamics and the energy of the Bose polaron for strong and repulsive interactions, complimenting previous spectroscopic measurements~\cite{jorgensen2016,ardila2019}.

The experiment is performed with $^{39}\text{K}$ atoms in an optical dipole trap~\cite{wacker2015} and follows the same procedure as the investigations for attractive impurity-medium interactions~\cite{Skou2020}. The medium is a BEC in the $\ket{F=1,m_F=-1}\equiv \ket{1}$ hyperfine state, where $F$ and $m_F$ denote the total angular momentum quantum number and its projection, respectively. The $\ket{F=1,m_F=0}\equiv \ket{2}$ hyperfine state constitutes the impurity state. The interaction between atoms in these two states is characterized by the scattering length, $a$, which can be controlled with a magnetic Fesh\-bach resonance at $113.8 G$~\cite{lysebo2010,tanzi2018}. As a result, the interaction strength $k_n a$ can be tuned at will from attractive, $k_n a < 0$, to repulsive, $k_n a > 0$, interactions, where $k_n = (6\pi^2 n_B)^{1/3}$ is determined by the average density, $n_B$, of the BEC. The typical energy scale of the system is given by $E_n = \hslash^2 k_n^2 / (2m)$, with the mass $m$ of ${}^{39}K$, which also provides the characteristic timescale $t_n = \hslash/E_n$. 

The interferometric sequence is initiated by a radio frequency~(rf) pulse on resonance with the atomic transition between the $\ket{1}$ and $\ket{2}$ states~\cite{fletcher2017}, which produces a small~$5\%$ coherent admixture of the impurity state. Subsequently, the system evolves at the chosen interaction strength $k_n a$ for a variable time, $t$, until a second rf pulse with a variable phase $\phi$ between $-\pi$ and $\pi$ is applied. We extract the signal from the loss of atoms due to three-body recombination, involving two condensate atoms and one impurity. The final number of atoms in the BEC has a sinusoidal dependence on the phase of the second rf pulse, which we parametrize as  $N(\phi) = N_0 - \mathcal{A}\cos(\phi - \phi_c)$. The normalized coherence function is then given by, $C(t) = \left|\mathcal{A}(t)/\mathcal{A}(0)\right| e^{i \phi_c(t)} $, which corresponds to the time-dependent Green's function of the impurity, $G(t) = -iC(t) = -i\bra{\Psi_{\text{BEC}}} \hat{c}(t)\hat{c}^\dagger(0) \ket{\Psi_{\text{BEC}}}$, with the state of the BEC $\ket{\Psi_{\text{BEC}}}$, and the creation operator for the impurity, $\hat{c}^\dagger$~\cite{cetina2016,Skou2020}. 

\begin{figure*}[t!]
	\begin{center}
		\includegraphics[width=1\textwidth]{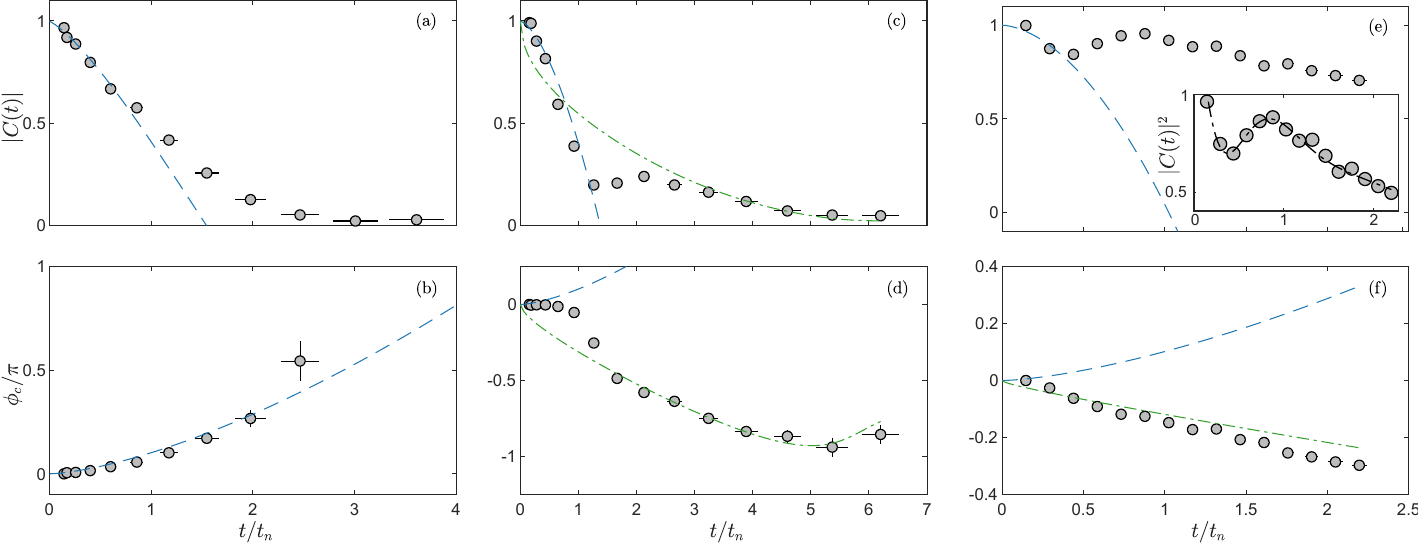}
	\end{center}
	\caption{Impurity dynamics at three characteristic interaction strengths. The coherence function is given for (a,b) strong $1/(k_na) = 0.4$, (c,d) intermediate $1/(k_na) = 0.7$, and (c,f) weak $1/(k_na) = 1.5$ interaction strengths. For comparison, theoretical results for the unitarity-limited case in Eq.~\eqref{Eq: unitarity} (blue dashed line) and for weak two-body interaction Eq.~\eqref{Eq: coherence_weak} (green dash-dotted line) are shown, without any free fitting parameters. Additional decoherence effects in the system due to three-body losses and the inhomogeneous density distribution, beyond Eqs.~\eqref{Eq: unitarity} and~\eqref{Eq: coherence_weak}, are accounted for in the theoretical description~\cite{SupplementaryMaterial}.}
	\label{fig:measurement}
\end{figure*} 

The observed coherence amplitude and phase evolution are shown in Fig.~\ref{fig:measurement} for three interaction strengths on the repulsive side of the resonance, displaying a number of striking features. The coherence amplitude shows a fast initial decay and a clear revival for intermediate and weak interaction strengths in Figs.~\ref{fig:measurement}~(c) and~\ref{fig:measurement}~(e). The phase initially increases and then reverses on a characteristic timescale, in Fig.~\ref{fig:measurement}~(d), indicating the crossover between different dynamical regimes. 

Importantly, the dynamics entails the population of a continuum of states, which together with decoherence effects in the system, leads to a decrease in the coherence amplitude. On top of this, the energy difference between the attractive and repulsive branches (see Fig.~\ref{fig:ind_fig}) leads to a quantum beat signal. In the following, the results shown in Fig.~\ref{fig:measurement} are discussed in detail, and the energies of the attractive and repulsive branches are inferred from the signal.

Theoretically, the coherence function can be obtained from a Fourier transform of the spectral function. At short times, an expansion in orders of $t/t_n$ yields
\begin{align}
C(t) &= 1 - (1 - i)k_{3/2}\left(\frac{t}{t_n}\right)^{3/2} - k_2\left( \frac{t}{t_n} \right)^2,
\label{Eq: unitarity}
\end{align}
valid for times $t\ll t_a=ma^2/\hslash$.  The second term is a consequence of unitarity-limited two-body interactions~\cite{braaten2010} with the universal constant $k_{3/2} = 16/(9\pi^{3/2})$~\cite{Skou2020,Parish2016}. The third term is an approximate second-order correction with $k_2 = \left( Z_g \left(E_g/E_n\right)^2/2 - 4/(3\pi k_n|a|) \right)$, which takes the attractive polaron with energy $E_g$ and residue $Z_g$, as well as the high-frequency tail of the spectral function~\cite{Skou2022PRR} into account.

The measured coherence function in Figs.~\ref{fig:measurement}~(a) and~\ref{fig:measurement}~(b) for strong repulsive interactions $1/(k_na) =0.4$ indeed agrees well with Eq.~\eqref{Eq: unitarity} at short times. We are, thus, able to quantitatively account for both the fast initial decay of the coherence amplitude as well as the initial positive phase evolution~\cite{footnote1}. At intermediate interaction strengths of $1/(k_na) =0.7$ in Figs.~\ref{fig:measurement}~(c) and~\ref{fig:measurement}~(d), similarly good agreement is obtained for short times, showing the presence of a universal regime, where Eq.~\eqref{Eq: unitarity} is valid, also in this case. For weak interactions at $1/(k_na)=1.5$, shown in Figs.~\ref{fig:measurement}~(e) and~\ref{fig:measurement}~(f), the universal regime contains only a few data points, since $t_a = 0.2t_n$.

For intermediate to weak interactions, the attractive polaron state lies close to the molecular state, whose energy to first order is given by $-\hslash^2/ma^2$. Therefore, the quantum beat signal between the two branches enters the dynamics on the same timescale, $t_a = ma^2 / \hslash$, as the system exits the unitarity-limited regime~\cite{SupplementaryMaterial}. As a result, we expect a dynamical regime described by weak coupling dynamics superimposed with quantum beats after $t = t_a$, which is qualitatively different from the case of attractive interactions, $1/(k_na) < 0$, where no beating is observed. The weak coupling dynamics is to a good approximation described by
\begin{align}
\begin{split}
C(t) &= e^{-iE_{\text{mf}}t/\hslash}e^{-(1+i)(t/t_w)^{1/2}}.
\end{split}
\label{Eq: coherence_weak}
\end{align}
In this case the dynamics is governed by the mean-field phase evolution with $E_{\rm mf} = \frac{4\pi \hslash^2 a n_B}{m}$, and the coherence decays on an interaction strength dependent timescale  $t_w = \frac{m}{32\pi\hslash n_B^2a^4}$~\cite{nielsen2019}, following a stretched-exponential form. 

For longer times, the measured coherence amplitude and phase in Figs.~\ref{fig:measurement}~(c) and~\ref{fig:measurement}~(d) both agree well with this prediction, without, however, capturing the crossover from the initial dynamics or the beat signal. This is expected since Eq.~\eqref{Eq: coherence_weak} does not include the low-lying attractive polaron branch.

\begin{figure}[ht]
    \centering
    \includegraphics[width=\columnwidth]{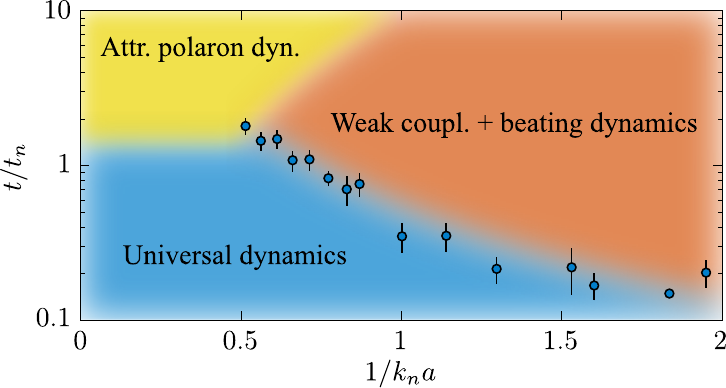}
    \caption{Dynamics of impurity evolution for repulsive interactions. The characteristic dynamical regimes are indicated as a function of inverse interaction strength $1/(k_n a)$ and evolution time $t/t_n$. The experimentally extracted crossover times (blue dots) indicate the transition from universal (blue region) to weak coupling and beating dynamics (orange region). For strong interactions and long times, the repulsive polaron decays, and the system is expected to transition into another dynamical regime (yellow region), in which the attractive polaron dominates the dynamics.}
    \label{fig:phase_diagram}
\end{figure}

To examine the crossover in detail, the associated crossover time is extracted from the coherence phase~\cite{something}. We pinpoint the time at which the experimentally extracted coherence phase becomes closer to the result of Eq.~\eqref{Eq: coherence_weak} rather than Eq.~\eqref{Eq: unitarity}. The crossover time is identified with the time between this and the previous data point. The extracted times are presented in Fig.~\ref{fig:phase_diagram}, along with the expected dynamical regimes. The extracted crossover timescale is in very good agreement with the transition between unitarity-limited dynamics (blue region) and weak-coupling beating dynamics (orange region), set by the timescale $t_a = ma^2 / \hslash$ for $1/(k_na) > 0.5$. This shows the existence of universal behavior at short times for all interaction strengths, extending the result in Ref.~\cite{Skou2020} to repulsive interactions. 

For strong interactions, there is no regime where Eq.~\eqref{Eq: coherence_weak} is accurate. Instead, the short-time unitarity-limited dynamics given by Eq.~\eqref{Eq: unitarity} is expected to transition directly to many-body dynamics for later times, consistent with the case of strong attractive interactions~\cite{Skou2020}. Theoretically, it is expected that the repulsive polaron eventually dampens out, resulting in the attractive polaron dominating the dynamics, at which point the system enters a new dynamical regime (Fig.~\ref{fig:phase_diagram}~yellow region). Experimentally, however, we did not observe signatures of this crossover. This is expected, since we theoretically observe that trap-induced density inhomogeneities significantly delay the crossover~\cite{SupplementaryMaterial}.

We now turn to analyze the quantum beat signal, clearly visible in Figs.~\ref{fig:measurement}~(c) and~\ref{fig:measurement}~(e), and present throughout the orange region in Fig.~\ref{fig:phase_diagram}. Importantly, such a signal is not present for attractive interactions of $1/(k_na) < 0$, but a similar feature was observed in interferometric investigations of the Fermi polaron~\cite{cetina2016}. In close analogy, the signal stems from the quantum interference of the repulsive and attractive branches, as demonstrated in the following. We apply a phenomenological model for the coherence function, 
\begin{align}
    C(t) = Z_g \, e^{-iE_g t/\hslash} + Z_r \, e^{-iE_r t/\hslash}e^{- \Gamma_r t/2}. \label{eq:two_state}
\end{align}
This describes two states; a ground state at energy $E_g$, and a damped excited state with energy $E_r > E_g$ and damping rate $\Gamma_r$, corresponding to the attractive and the repulsive polarons with residues $Z_g$ and $Z_r$, respectively. This implies two main features of the coherence function for weak-coupling beating dynamics. The slope of the phase corresponds to the energy of the repulsive branch since it is expected to have the largest residue, i.e.\ $\phi_C\approx-E_r t/\hslash$. In addition, the squared amplitude of the coherence function contains a sinusoidal oscillation at a frequency corresponding to the energy difference between the two branches.

Based on this approach, the experimental coherence amplitude squared and phase are fitted with the functions
\begin{align}
|\mathcal{C}(t)|^2 &= A\, e^{-t/\tau_1} \cos\omega t+B\,  e^{-t/\tau_2},\label{Eq:fit_amplitude}\\
\varphi_c &= A_\varphi - B_\varphi\, t. \label{Eq: fit_phase}
\end{align}
This corresponds to the modulus squared of Eq.~\eqref{eq:two_state} with an additional exponential decay of the oscillation amplitude to model three-body decay and other loss mechanisms. The energy difference can be extracted from the fitting parameters as $E_r - E_g = \hslash\omega$. An example of the coherence amplitude fit is shown in the inset of Fig.~\ref{fig:measurement}~(e). Generally, a single beat is resolved, due to the decoherence effects and losses in the experiment discussed above~\cite{SupplementaryMaterial}. Furthermore, the repulsive branch is expected to be damped, such that the beat signal, even in an ideal setting, attenuates in time. 

\begin{figure}[t!]
    \centering
    \includegraphics[width=\columnwidth]{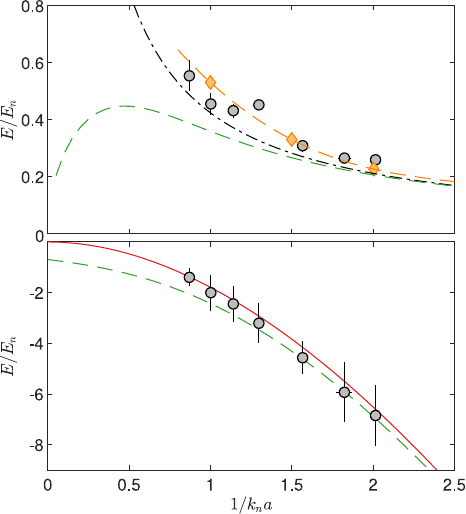}
    \caption{Energies of the repulsive and attractive branches as a function of the inverse interaction strength. The measured energies are shown as gray points and the result of our diagrammatic calculation is indicated by green dashed lines~\cite{SupplementaryMaterial}. (top panel) The repulsive branch energies lie clearly above the simple mean-field expectation (black dash-dotted line) and agree with a quantum Monte-Carlo prediction~\cite{ardila2019} (yellow dashed line, diamonds). (bottom panel) The attractive branch energies do not allow for a distinction between the diagrammatic result and the molecular energy (red line).}
    \label{fig:Energy_figure}
\end{figure}

Additionally, the fit directly yields the energy of the repulsive branch, $E_r = \hslash B_\varphi$. This technique, however, can only be used when the phase evolution shows a linear evolution for long times. This is the case for the phase evolution in Fig.~\ref{fig:measurement}~(f) but not for Fig.~\ref{fig:measurement}~(d), where the last data point shows an increase. This limits our analysis to $1/(k_na) > 0.85$. To avoid the influence of the unitary regime, the linear fit is, additionally, restricted to $t > t_a$. Figure~\ref{fig:Energy_figure}~(upper panel) shows the obtained energies for the repulsive branch for this range of interaction strengths. These energies lie consistently higher than the simple mean-field expectation and agree with a Monte-Carlo prediction~\cite{ardila2019}. Moreover, they lie systematically higher than our diagrammatic calculation based on the ladder approximation~\cite{SupplementaryMaterial}.

Based on these results, the energy of the attractive branch is obtained by subtracting the extracted energy difference from the repulsive branch energy $E_g =E_r -\hslash\omega$ as shown in Fig.~\ref{fig:Energy_figure}~(lower panel). The energies agree with both the attractive polaron energy, obtained from diagrammatic calculations, and the molecular energy with an effective range given by the Van der Waals length of our medium, $R^*=60a_0$~\cite{jorgensen2016,SupplementaryMaterial}. Experimentally, we cannot distinguish between these two results. However, the rf response of the molecular state is expected to be very small compared to the attractive polaron state, as has been previously measured for the Fermi polaron~\cite{Kohstall2012}. In particular, since the molecular state does not have a plane wave component, we expect such a signal to scale with the density of generated impurities. This is indeed much weaker than the response from the attractive polaron state scaling with the density of the entire BEC. With this in mind, the above agreement confirms our interpretation of the beat signal arising from an interference between the two polaron states. Furthermore, the simple linear fitting for the repulsive branch energy yields a robust measurement beyond the accuracy obtained spectroscopically~\cite{jorgensen2016}. Note that we see no evidence of the recently predicted presence of meta-stable states at intermediate energies between the two branches~\cite{mostaan2023unified}.

In conclusion, we have performed Ramsey interferometry on a BEC of ${}^{39}$K to investigate the physics of Bose polarons in a wide range of repulsive interactions. The observed quantum beat signal is consistent with a repulsive and attractive quasiparticle branch that form and exist in a coherent superposition. From the beat frequency and the slope of the coherence phase, we determine the energy of these branches. The repulsive branch lies systematically above the simple mean-field expectation and agrees well with quantum Monte-Carlo calculations~\cite{ardila2019}. Furthermore, the initial dynamics is seen to agree very well with the exact unitarity-limited behavior with the non-analytical time-dependence of $t^{3/2}$~\cite{Skou2020,Parish2016}, and the crossover to a weak-coupling regime closely follows the expected $t = ma^2 / \hslash$ behavior for $1/(k_na) > 0.5$. At even stronger interactions of $1/(k_na) < 0.5$, the dynamics follows the unitarity-limited behavior for long times. This, along with the quantitative discrepancy from the diagrammatic calculations, calls for further theoretical analyses, especially at intermediate to strong repulsive interactions of $0< 1/(k_na) < 1$. \\

\begin{acknowledgments}
The authors thank Ragheed Alhyder for valuable discussions and Adam Chatterley for the careful review of the manuscript. This work was supported by the Danish National Research Foundation through the Center of Excellence “CCQ” (Grant no. DNRF156), by the Novo Nordisk Foundation NERD grant (Grant no. NNF22OC0075986), and by the Independent Research Fund Denmark (Grant no. 0135-00205B). KKN acknowledges support from the Carlsberg Foundation through a Carlsberg Internationalisation Fellowship. 
\end{acknowledgments}

\bibliography{ms}

\begin{thebibliography}{43}
\bibitem [{foo()}]{footnote1}%
  \BibitemOpen
  \href@noop {} {}\bibinfo {note} {Interestingly, we find that this behaviour
  of the phase persists for relatively long times beyond $t \simeq 1.4t_n$ for
  $1/(k_na) \lesssim 0.5$ as exemplified in Fig.~\ref{fig:measurement}(b),
  where a direct crossover to the many-body regime was observed for attractive
  interactions~\cite{Skou2020}.}\BibitemShut {Stop}%
\end{thebibliography}
 
\end{document}


\beginsupplement

\title{SUPPLEMENTARY INFORMATION FOR “QUANTUM BEAT SPECTROSCOPY OF REPULSIVE BOSE POLARONS"}

\author{A. \ M. \ Morgen}
\affiliation{Center for Complex Quantum Systems, Department of Physics and Astronomy, Aarhus University, Ny Munkegade 120, DK-8000 Aarhus C, Denmark}
\author{S.\ S.\ Balling}
\affiliation{Center for Complex Quantum Systems, Department of Physics and Astronomy, Aarhus University, Ny Munkegade 120, DK-8000 Aarhus C, Denmark}
\author{K.\ Knakkergaard Nielsen}
\affiliation{Max-Planck Institute for Quantum Optics, Hans-Kopfermann-Str. 1, D-85748 Garching, Germany}
\author{T.\ Pohl}
\affiliation{Center for Complex Quantum Systems, Department of Physics and Astronomy, Aarhus University, Ny Munkegade 120, DK-8000 Aarhus C, Denmark}
\affiliation{Institute for Theoretical Physics, Vienna University of Technology, Wiedner Hauptstraße 8-10, 1040 Vienna, Austria}
\author{G.\ M.\ Bruun}
\affiliation{Center for Complex Quantum Systems, Department of Physics and Astronomy, Aarhus University, Ny Munkegade 120, DK-8000 Aarhus C, Denmark}
\author{J. J. Arlt}
\affiliation{Center for Complex Quantum Systems, Department of Physics and Astronomy, Aarhus University, Ny Munkegade 120, DK-8000 Aarhus C, Denmark}
\date{\today}

\maketitle
\section{Experimental Data Analysis}

\subsection{Extracted energy differences}
The quantum beat signal observed for the coherence amplitude in the main manuscript was fitted with a phenomenological model and the energy difference between the repulsive and attractive branches was extracted. In combination with  the linear fit of the phase evolution, this was used to find the energy of the attractive and repulsive branches. 

\begin{figure}[h!]
    \centering
    \includegraphics[width=0.5\columnwidth]{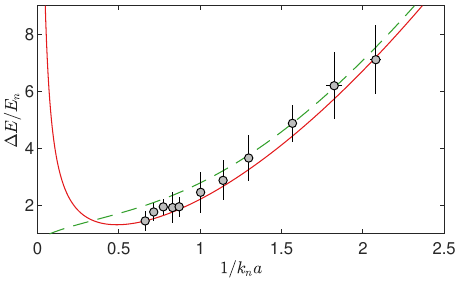}
    \caption{\textbf{Energy differences.} Energy differences between the attractive and repulsive branches extracted from the beat signal in the coherence amplitude (gray points). For comparison, the energy difference between the attractive and repulsive polaron energy obtained from the diagrammatic calculation in Sec.~\ref{sec:PolEnergy} (green dashed line) is shown. Moreover, the energy difference between the mean-field energy and the molecular energy (red line) is indicated.}
    \label{fig:Ediffs}
\end{figure}

Figure~\ref{fig:Ediffs} shows all extracted energy differences as a function of the inverse interaction strength. Additional energy differences at high interaction strengths are included, where the linear fit of the phase evolution fails but a quantum beat is still observed in the coherence amplitude, as is the case for $1/k_na = 0.7$ in Fig.~2~(c) and~(d) in the main manuscript. The extracted energy differences show a general increase from strong to weak  interactions, which is primarily due to the large negative energy of the attractive branch. For large interaction strengths and consequently small values of $1/(k_n a)<0.7$, no quantum beat signals are observed. We attribute this to the rapid damping of the repulsive polaron along with the universal decay of the coherence function and rapid decoherence processes in our experiment. 

Figure~\ref{fig:Ediffs} also compares the data with, the energy difference between the attractive and repulsive polaron energy calculated in Sec.~\ref{sec:PolEnergy} and with the energy difference between the mean-field energy and the molecular energy. The data is in good agreement with both predictions but does not allow for a distinction between them.

\subsection{Decoherence from impurity losses}
\begin{figure}[h!]
    \centering
    \includegraphics[width=0.5\columnwidth]{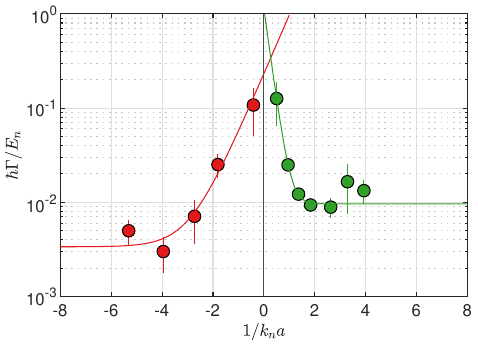}
    \caption{\textbf{Loss rates}. Measured loss rates of impurity atoms embedded in a Bose-Einstein condensate as a function of the inverse interaction strength. The measured loss rates for repulsive interactions (green points) are shown together with previous experimental data for attractive interactions~\cite{Skou2020} (red points). The solid lines indicate fits to the data (see text).}
    \label{fig:loss_rep}
\end{figure}

The loss of impurity atoms leads to faster decoherence of the system and is mainly caused by three-body losses with medium atoms, which has to be accounted for. The lifetime of impurity atoms was previously investigated for attractive interactions between the impurity and the medium~\cite{Skou2020} and the same approach is used here for repulsive interactions. In brief, the experimental sequence starts with a Bose-Einstein condensate (BEC) in the $\ket{F=1,m_F=-1}$ medium state. An initial radio frequency pulse produces a $10\%$ admixture in the $\ket{F=1,m_F=0}$ impurity state. In the following evolution, impurity atoms are predominantly lost due to three body collisions, where two medium atoms are lost for each impurity atom. After a variable wait time the remaining impurities are transferred to the $\ket{F=1,m_F=1}$ state using a $\pi$-pulse. In this state the impurity atoms are quickly lost due to two-body collisions with $\ket{F=1,m_F=-1}$ atoms. The final atom number in the $\ket{F=1,m_F=-1}$ state is then observed as a function of the waiting time for various interaction strengths. Since the two loss processes lead to a different number of lost medium atoms, this sequence allows for a measurement of the impurity lifetime by measuring the number of medium atoms. An exponential fit of the form $\sim \exp(-\Gamma_{\text{loss}}t)$ is applied for each data set, to extract the loss rate $\Gamma_{\text{loss}}$. The observed loss rates for repulsive interactions (green points) are shown as a function of inverse interaction strength in Fig.~\ref{fig:loss_rep}. For comparison, Fig.~\ref{fig:loss_rep} also includes the  previously measured loss rates for attractive interactions (red points)~\cite{Skou2020}. An empirical fit of the form, $\beta_1 + \beta_2 \exp(\beta_3/k_na)$ was performed to extract the loss rate at arbitrary interaction strengths. The theoretically calculated coherence function is then modified according to $C(t) \rightarrow C(t)\exp(-\Gamma_{\text{loss}}t)$, before comparison with the experimental results in Fig.~2 of the main manuscript.

\subsection{Decoherence from density inhomogeneity}
\label{sec:DecInhom}
The harmonic trap used in the experiment leads to an inhomogeneous density distribution of the atomic clouds. This causes additional decoherence of the system since the impurities at high medium density at the center of the cloud evolve differently compared to those at the edge. 

It is therefore necessary to perform a local density approximation of the theoretical coherence function~\cite{Skou2020}. This corresponds to an integration of the coherence function over its density dependent terms according to
\begin{align}
\langle C(t) \rangle = \frac{15}{4n_0}\int_{0}^{n_0}dn \frac{n}{n_0}\sqrt{1-\frac{n}{n_0}}C(t),
\end{align}
in the Thomas-Fermi approximation. The theoretically calculated coherence function is then modified according to $C(t) \rightarrow \langle C(t) \rangle$, for comparison with the experimental results in Fig.~2 of the main manuscript.

\section{Theoretical Analysis}
\subsection{Exact short-time dynamics at weak coupling}
\begin{figure}[h!]
    \centering
    \includegraphics[width=0.8\columnwidth]{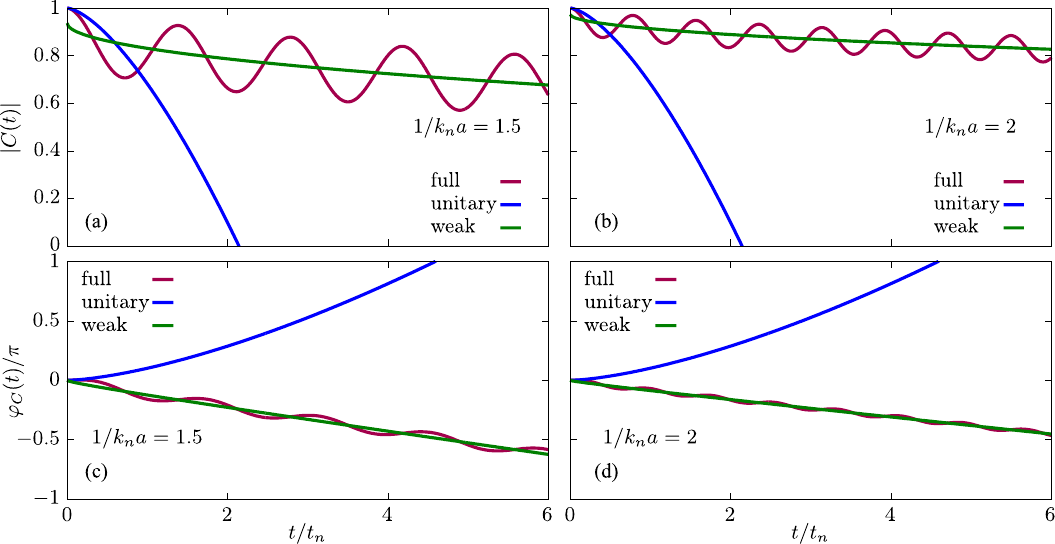}
    \caption{\textbf{Short-time dynamics at weak coupling}. The coherence amplitude (a), (b) and phase (c), (d) at indicated interaction strengths. The blue line describes unitary behavior valid for $t \ll t_a$ (Eq. \eqref{eq.unitary_regime}), the green line is described by Eq. \eqref{eq.coherence_weak}, while the full dynamics in red is described by Eq. \eqref{eq.coherence_short_time}.}
    \label{fig.short_time_dynamics_supplementary}
\end{figure}

Here, we derive the exact short-time dynamics for repulsive interactions at weak coupling $1 / k_na \gg 1$. The analysis is similar to the one described in the Supplementary Material of ref.~\cite{Skou2020}. The coherence dynamics is in general calculated from the Fourier transformation of the zero momentum spectral function, $C(t) = \int_{-\infty}^{+\infty} d\omega e^{-i\omega t} A(\omega) / 2\pi$. We may decompose this as
\begin{align}
C(t) &= \int_{-\infty}^{+\infty} \frac{d\omega}{2\pi} e^{-i\omega t} A(\omega) \nonumber \\
&=  \int_{-\infty}^{+\infty} \frac{d\omega}{2\pi} (1 - i\omega t) A(\omega) + \int_{0}^{+\infty} \frac{d\omega}{2\pi} \left[e^{-i\omega t} - (1 - i\omega t)\right] A(\omega) + \int_{-\infty}^{0} \frac{d\omega}{2\pi} \left[e^{-i\omega t} - (1 - i\omega t)\right] A(\omega). 
\label{eq.coherence_function_decomposition}
\end{align}
In refs.~\cite{braaten2010,Skou2020} it is shown that $\int_{-\infty}^{+\infty} \frac{d\omega}{2\pi} \omega A(\omega) \propto 4\pi \hslash a_B / m$, where $a_B$ is the medium-medium scattering length. Hence, this term can safely be omitted. The second term was determined from the exact high-frequency tail of the spectral function~\cite{braaten2010} to yield the contribution
\begin{equation}
C_{\rm tail}(t) = \int_{0}^{+\infty} \frac{d\omega}{2\pi} \left[e^{-i\omega t} - (1 - i\omega t)\right] A(\omega) = \frac{2}{3\pi} (k_n|a|)^3 \left[ 1 + \frac{it}{t_a} - \frac{2}{\sqrt{\pi}} \Gamma\left(\frac{3}{2}, \frac{it}{t_a}\right) \right],
\end{equation}
with $t_a = ma^2 / \hslash$, and $\Gamma(\nu, x) = \int_x^\infty ds\, s^{\nu - 1} e^s$ the upper incomplete gamma function. The third term in Eq. \eqref{eq.coherence_function_decomposition} can be omitted only at times shorter than $\hslash / |E_g|$, where $E_g$ is the energy of the attractive polaron. For $1/k_na \ll -1$, this coincides with the mean field timescale, whereas at unitarity $E_g \simeq -E_n$, such that $C(t) = 1 + C_{\rm tail}(t)$ up to times of order $t_n$. For positive scattering lengths, the attractive polaron, however, eventually approaches the energy $E_m = -\hslash^2 / ma^2$ of a universal Feshbach molecule. Since the associated timescale of this state coincides with $t_a$, $\hslash / |E_m| = t_a$, it will have a considerable contribution for times $t > t_a$. At weak coupling, this may be evaluated exactly, because the ladder approximation in this limit provides the correct perturbative description. In particular, as the energy approaches $E_m = -\hslash^2 / ma^2$, the residue can be computed from the expression~\cite{Yan2020}
\begin{equation}
Z_g = \frac{1}{1 + \frac{3\pi}{8\sqrt{2}}\left(\frac{|E_g|}{E_n}\right)^{3/2}} \to \frac{1}{1 + \frac{3\pi}{4}\frac{1}{(k_na)^3}} \to \frac{4}{3\pi}(k_na)^3. 
\end{equation}
In this regime, $A(\omega) = 2\pi Z_g \delta(\omega - E_m / \hslash)$ for $\omega < 0$. As a result, the third term in Eq. \eqref{eq.coherence_function_decomposition} at weak coupling is given by
\begin{align}
C_g(t) = \int_{-\infty}^{0} \frac{d\omega}{2\pi} \left[e^{-i\omega t} - (1 - i\omega t)\right] A(\omega) \to \frac{4}{3\pi}(k_na)^3 \left[e^{it / t_a} - \left(1 + \frac{it}{t_a}\right)\right]
\end{align}
The total short-time coherence function is then
\begin{align}
C(t) = 1 + C_{\rm tail}(t) + C_g(t) = 1 - \frac{2}{3\pi} (k_na)^3 \left[1 + \frac{it}{t_a} + 2 e^{it/t_a}\left(\frac{1}{\sqrt{\pi}}\Gamma\left(\frac{3}{2}, \frac{it}{t_a}\right) - 1\right)\right].
\label{eq.coherence_short_time}
\end{align}
For $t \ll t_a$, this yields the same unitarity-limited behavior as for negative scattering lengths, 
\begin{align}
C(t) = 1 - (1-i)k_{3/2}\left(\frac{t}{t_n}\right)^{3/2},
\label{eq.unitary_regime}
\end{align}
with $k_{3/2} = 16/(9\pi^{3/2})$. On the other hand, the dynamics for times $t > t_a$ is now qualitatively different from the one found for negative scattering lengths. In particular, comparing Eq. \eqref{eq.coherence_short_time} to the weak coupling behavior
\begin{align}
C_w(t) = 1 - \frac{2}{3\pi} (k_na)^3 - \frac{iE_{\rm mf}t}{\hslash} - (1+i)\sqrt{\frac{t}{t_w}},
\label{eq.coherence_weak}
\end{align}
with $t_w = \frac{m}{32\pi\hslash n_B^2a^4}$ in Fig. \ref{fig.short_time_dynamics_supplementary} demonstrates that the full dynamics shows oscillations around the behavior described by Eq. \eqref{eq.coherence_weak} at a frequency given by $1 / t_a = \hslash / ma^2$. The modification of the coherence phase is, however, seen to be minor.  

\subsection{Polaron energies in the ladder approximation}
\label{sec:PolEnergy}
Here, we briefly outline the calculation of the attractive and repulsive polaron energies in the ladder approximation~\cite{Rath2013}. In the experiment, the medium-medium scattering length is $k_na_B \simeq 0.01$. We, therefore, assume that the relevant physics can be explained by assuming an ideal BEC. The impurity Green's function $G^{-1}(\omega) = \omega + i\eta - \Sigma(\omega)$ (evaluated at zero momentum and equal masses), is expressed in terms of the self-energy $\Sigma(\omega)$, which in the ladder approximation may additionally be written as $\Sigma(\omega) = n_B \mathcal{T}(\omega)$. Here, 
\begin{align}
    \mathcal{T}(\omega) = \frac{1}{\frac{m}{4\pi\hslash^2\tilde{a}(\omega)} - \Pi(\omega)}
\end{align}
is the $\mathcal{T}$-matrix, and $1/\tilde{a}(\omega) = 1 / a + R^* m \omega$ includes the effect of the finite range parameter $R^*$~\cite{Fritsche2021}. For the Feshbach resonance in question the range parameter is estimated to be $R^* = 60a_0$~\cite{jorgensen2016}. We find this to be important to quantitatively capture the attractive polaron energy in the regime $1/k_na > 0.5$. Finally, $\Pi(\omega) = -im^{3/2}(\omega + i\eta)^{1/2} / (4\pi \hslash^{5/2})$ is the pair propagator, and $\eta$ is a positive infinitesimal. The attractive polaron energy is then determined by the equation $E_g / \hslash = \Sigma(E_g / \hslash)$, while the repulsive polaron energy satisfies $E_r / \hslash = {\rm Re}[\Sigma(E_r / \hslash)]$ for $E_r > 0$. We find that these equations can equivalently be expressed as
\begin{align}
\frac{E_g}{E_n} &= \frac{4}{3\pi}\frac{1}{\frac{1}{k_na} + k_n R^* \frac{E_g}{2E_n} - \sqrt{\frac{|E_g|}{2E_n}}}, \nonumber \\
\frac{E_r}{E_n} &= \frac{4}{3\pi}\frac{\frac{1}{k_na} + k_n R^* \frac{E_g}{2E_n}}{\left(\frac{1}{k_na} + k_n R^* \frac{E_r}{2E_n}\right)^2 + \frac{E_r}{2E_n}}.
\end{align}
Using $k_n R^* \simeq 0.06$, the numerical solution of these equations leads to the results in Fig. 4 of the main text. 

\subsection{Phase evolution with and without trap averaging}
Using the ladder approximation, we, here, provide strong theoretical evidence that the initial evolution of the coherence phase follows the repulsive polaron energy. We also show that trap averaging leads to a severe delay or even prevention of the crossover into the dynamical regime dominated by attractive polaron dynamics. \\

\begin{figure}[H]
    \centering
    \includegraphics[width=0.8\columnwidth]{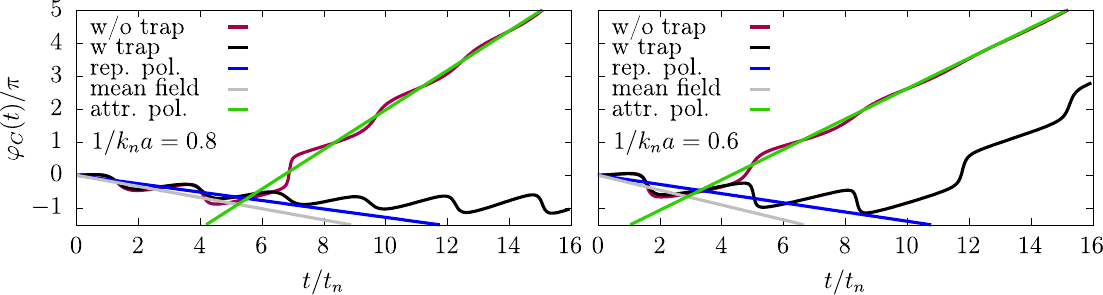}
    \caption{\textbf{Phase evolution in the ladder approximation}. The coherence phase is shown for two indicated interaction strengths for a homogeneous system (red line) as well as in a harmonic trap (black line). We also show the phase evolution corresponding to repulsive and attractive polarons (blue and green lines, respectively), as well as the mean field phase evolution in grey.}
    \label{fig.phase_ladder_approximation}
\end{figure}

The coherence function is calculated as the Fourier transform of the zero momentum spectral function, $C(t) = \int_{-\infty}^{+\infty} d\omega e^{-i\omega t} A(\omega) / 2\pi$. The spectral function $A(\omega) = -2{\rm Im} G(\omega)$ is evaluated within the ladder approximation described in the previous subsection. In particular, $G^{-1}(\omega) = \omega +i\eta - \Sigma(\omega)$, is evaluated from the self-energy
\begin{align}
\frac{\Sigma(\omega)}{E_n} = \frac{4/3\pi}{\frac{1}{k_na} + k_nR^* \frac{\hslash\omega}{2E_n} + i\sqrt{\frac{\hslash\omega}{2E_n}}}. 
\end{align}
In Fig. \ref{fig.phase_ladder_approximation}, we show the resulting phase evolution $\varphi_C(t) = \arg(C(t))$. This is shown both for a homogeneous system and in a harmonic trap. For the latter, we use a local density approximation to evaluate the trap-averaged coherence function $\langle C(t)\rangle = \frac{15}{4n_0}\int_{0}^{n_0}dn \frac{n}{n_0}\sqrt{1-\frac{n}{n_0}}C(n,t)$, as described previously in Sec.~\ref{sec:DecInhom}. The key insights of this calculations are the following. (1) The initial behavior of the phase follows the one corresponding to the repulsive polaron: $-iE_r t / \hslash$, and not the mean-field energy. This is also the case for the system in a trap. Experimentally, this allows us to extract the repulsive polaron energy from the short-time linear slope of the phase. (2) For a homogeneous system, as the repulsive polaron dampens out, the phase rapidly switches behavior and starts following the attractive polaron phase evolution $-iE_g t / \hslash$. (3) In the harmonic trap, this crossover is either severely delayed as in Fig. \ref{fig.phase_ladder_approximation}(right), or entirely washed out as in Fig. \ref{fig.phase_ladder_approximation}(left).\\

We find, however, that the ladder approximation does not quantitatively describe the phase evolution. It seems to overestimate the overall amplitude of the beat signal between the repulsive and attractive polarons. Furthermore, the repulsive polaron energies extracted in the experiment lie systematically above the ones found in the ladder approximation. 

\bibliography{supplement}